\documentclass{article}
\usepackage{amsmath}
\usepackage{graphicx}
\usepackage[dvipsnames]{xcolor}
\usepackage[bottom]{footmisc}

\title{Gain-through-filtering enables tuneable frequency comb generation in passive optical resonators}
\author{
Florent Bessin$^{1}$, Auro M. Perego$^{2,*}$, Kestutis Staliunas$^{3,4}$, \\ Sergei K. Turitsyn$^{2,5}$, Alexandre Kudlinski$^{1}$, Matteo Conforti,$^{1}$ \\ and Arnaud Mussot$^{1}$}

\begin{document}

\maketitle

 $^{1}$Univ. Lille, CNRS, UMR 8523-PhLAM Physique des Lasers Atomes et Mol\'ecules, F-59000 Lille, France\newline
 $^{2}$Aston Institute of Photonics Technologies, Aston University, Birmingham, B4 7ET, UK\newline
 $^{3}$Instituci\'{o} Catalana de Recerca i Estudis Avan\c{c}ats, Pg. Lluis Companys 23, 08010, Barcelona, Spain\newline
 $^{4}$Departament de F\'{i}sica, Universitat Polit\`{e}cnica de Catalunya, 08222, Terrassa, Spain\newline
 $^{5}$Novosibirsk State University, Novosibirsk, 630090, Russia \newline
\flushleft$^*$Corresponding author (a.perego1@aston.ac.uk)
\newline

\begin{abstract}
Optical frequency combs (OFCs), consisting of a set of phase locked equally spaced laser frequency lines, have enabled a great leap in precision spectroscopy and metrology since seminal works of H\"{a}nsch \textit{et al.}. Nowadays, OFCs are cornerstones of a wealth of further applications ranging from chemistry and biology to astrophysics and including molecular fingerprinting and LIDARs among others. Driven passive optical resonators constitute the ideal platform for OFCs generation in terms of compactness and low energy footprint. We propose here a new technique for generation of OFCs with tuneable repetition rate in externally driven optical resonators based on the gain-through-filtering process, a simple and elegant method, due to an asymmetric spectral filtering on one side of the pump wave. We demonstrate a proof-of-concept experimental result in a fibre resonator, pioneering a new technique that does not require specific engineering of the resonator dispersion to generate frequency agile OFCs.
\end{abstract}\newpage
\textbf{\large Introduction}

In 2005 the Nobel prize for physics was awarded to Hall and H\"{a}nsch for the discovery of optical frequency combs (OFCs) and their applications in ultraprecise spectroscopy measurements \cite{Hansch}. An outstanding effort of the research community is now aiming at developing OFCs based technology with an impressive range of applications in exoplanets detection \cite{Astro}, metrology \cite{Udem}, spectroscopy and molecular fingerprinting \cite{Coddington, fingerprinting} and ultraprecise distance measurements \cite{Coddington2} to mention just the most relevant ones.
OFCs can be generated by a variety of methods including mode-locked lasers \cite{Jones}, spontaneous four-wave mixing in quantum cascade laser \cite{Hugi}, difference frequency generation \cite{difference}, parametric four-wave mixing in highly nonlinear fibres \cite{HNLF} and externally driven passive resonators either with quadratic \cite{quadratic} nonlinearity and (mainly) with cubic Kerr nonlinearity in mircoresonators \cite{Kippenberg}. Microresonator technology attracts a special interest due to compactness of the device reaching up to chip dimensions which makes them extremely suitable for applications in the field \cite{DelHaye,DelHaye2,Chip}. Moreover, microresonator quality factor ranging from one million to one billion, leads to a strong confinement of the light circulating within the loop \cite{Vahala,Vahala2, Maleki}. Consequently, from relatively weak continuous wave (CW) lasers it is possible to generate broadband OFCs in these systems, spanning over more than one octave to obtain the $f-2f$ self-referencing scheme \cite{Udem}, or to generate OFCs in the visible range \cite{Vahala17} for chip scale photonic clocks \cite{Papp2014}, or in the infrared region \cite{Chip, Yu} that contains  strong and distinctive patterns of spectroscopic signatures, providing unique molecular fingerprints. To achieve such impressive spectral broadenings, temporal cavity solitons are utilised \cite{leocavsol,Lugiato,Chembo} and cavity solitons based OFCs have indeed been demonstrated experimentally in microresonators \cite{Herr2, Yi}. One of the important technical challenges here is that anomalous dispersion is required to generate these combs which is difficult to achieve due to practical limitations imposed by fabrication processes, specifically in the visible range of the spectrum where material dispersion dominates. Alternative options have been proposed to generate OFCs in normal dispersion Kerr resonators based on Faraday instability, which requires in addition a careful longitudinal dispersion management \cite{Wong}, or on dark cavity solitons \cite{Dark}. In addition, the repetition rate can not be tuned in these resonators since it is fixed by the resonator opto-geometrical parameters. 

In this work we propose a concept based on the gain-through-filtering (GTF) induced MI to achieve Kerr OFCs formation in normal dispersion driven optical resonators. GTF is a generalisation of the gain-through loss mechanism, a recently proposed parametric process that takes place when losses induced by an asymmetric spectral filter exhibit unequal strength for signal and idler waves \cite{Perego,Tanemura}. Such loss unbalance counter-intuitively drives energy transfer from a powerful pump field to two spectral sidebands, the first located in the vicinity of the maximum losses point (signal) and the second one (idler) symmetric with respect to the pump frequency. The mechanism has been extended to a broader context, highlighting its potential for various applications in photonics including on demand signal amplification, improved optical parametric oscillators and ultrafast generation of pulses \cite{Perego}, both in single waveguides and in coupled planar structures \cite{El}.  GTF generalises this concept by including the contribution of dispersion naturally associated to the filter dissipation profile through causality relation, thus providing an additional degree of freedom to harness the process. 
Here we demonstrate for the first time GTF induced OFCs generation in a passive optical fibre resonator with an intra-cavity spectral filter to trigger the amplification process. GTF induces first signal and idler amplification, which eventually leads to multiple sidebands generation through cascaded four-wave mixing. We show how GTF technique allows an easy tuneability of the comb repetition rate (comb teeth spacing), by simply varying the detuning between pump and loss frequency. Furthermore, to get a deeper physical insight, we derived a simple analytic expression that highlights the contribution of the filter onto the phase matching relation ruling the original MI process which initiates OFC formation. 

\begin{figure}[ht!]
\centering
\includegraphics[width=12cm,keepaspectratio]{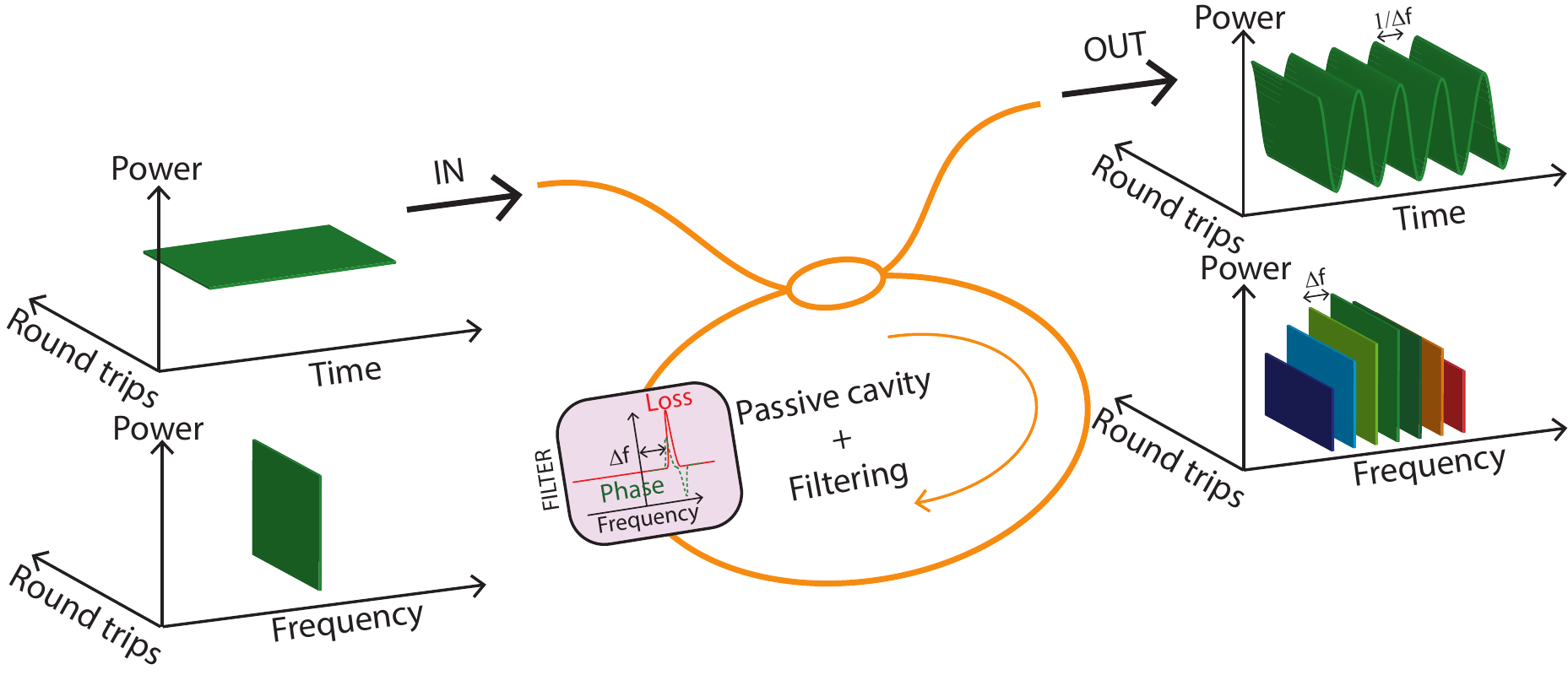}
\caption{Concept of GTF comb in a passive resonator. A fibre ring resonator, including an asymmetric loss element (filter) is pumped by a CW laser. At the output, an OFC is generated with a teeth spacing imposed by the frequency shift $\Delta f$ between the pump and the filter.}
\label{fig:1}
\end{figure}
\textbf{\large Results}

\begin{figure}[ht!]
\centering
\includegraphics[width=10cm,keepaspectratio]{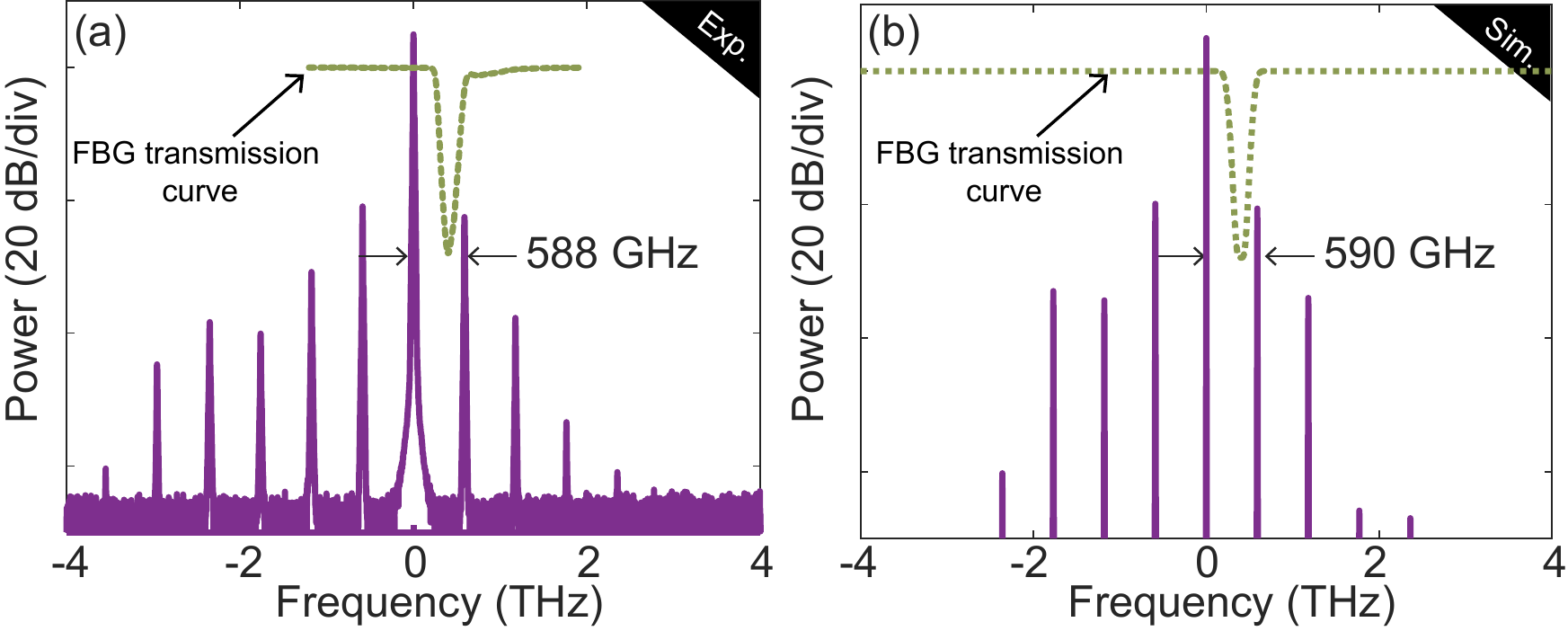}
\caption{GTF comb. 
Optical frequency comb at the cavity output for $\lambda_\text{P}=1544.66$ nm (a) in experiments and (b) in numerics. The green dashed lines represents the modulus square of the transfer function of the notch filter.}
\label{fig:2}
\end{figure}

\textbf{GTF frequency comb.} We built a fibre ring resonator from a 104.2 m long dispersion shifted fibre, in order to have normal dispersion, $\beta_2=+0.5$ ps$^2$/km, at the pump wavelength $\lambda_\text{P}$. A notch filter with 330 GHz full width at half maximum, made from a fibre Bragg grating used in transmission, leads to 28 dB loss at maximum, and is detuned by about $\Delta f$=400 GHz from the pump (see the cartoon in Fig. \ref{fig:1} and detailed experimental scheme in Supplementary Note 1 and Supplementary Fig. 1). Square profile pump pulses of 1.5 ns duration are launched in the cavity through a 90/10 coupler and the whole system is stabilised by means of an electronic feedback loop to keep both the driving frequency and the overall cavity length stable \cite{Coen98}. By pumping the resonator above the instability threshold with 6.6 W pump power, we observe the generation of an harmonic frequency comb with a teeth spacing of 588 GHz, as can be seen in Fig.\ref{fig:2} (a) which showcases the cavity output spectrum. More than 11 lines are generated due to subsequent four wave mixing processes between the pump, signal and idler waves originally triggered by the GTF process. For a purely dissipative filter the maximum of the GTF should occur at the maximally attenuated frequency \cite{Perego}. However, the causality principle imposes, through Kramers-Kronig relations, to consider the complex response function of the filter. As discussed in  detail in Supplementary Information (see Supplementary Note 3), the contribution of the phase of the filter transfer function leads to an additional frequency shift with respect to the one predicted by the basic gain-through loss theory.  Hence, rather than being located at the maximum loss position of the filter (400 GHz), the positions of the sidebands experience an extra frequency shift and are located at 588 GHz in this particular example (Fig. \ref{fig:2} (a)).  Numerical simulations performed by integrating the full Ikeda map equations (details of numerical simulations are given in Methods section) give results in good agreement with the experimental ones, as it can be appreciated from Fig. \ref{fig:2} (b).  

By developing a linear stability analysis on the Ikeda map model \cite{Coen1997}, we derived the following approximated relation highlighting the contribution of the filtering on the phase matching process (see Supplementary Note 3 for more details):
\begin{equation}\label{PM}
    \frac{\beta_2}{2}\omega^2L+ 2\gamma P L + \phi_0 + \psi_\text{e} (\omega)=0.
\end{equation}
Here, $\beta_2$ is the group velocity dispersion of the fibre, $\omega$ the pump-sidebands frequency shift, $\gamma$ the nonlinear coefficient of the fibre, $P$ the intra-cavity pump power, $L$ the fibre length, $ \phi_0$ the linear cavity phase shift and $\psi_e(\omega)$ the even part of the phase of the filter transfer function. From this relation, we found a frequency shift of 580 GHz, in excellent agreement with experiments and numerics giving 588 GHz and 590 GHz respectively. Output cavity spectra in experiments and in numerics are not symmetric with respect to the pump frequency, and a larger number of teeth are generated on the Stokes side. We have numerically checked that this asymmetry originates from the dispersion slope (due to contributions from both fibre and filter), as it has already been reported in \cite{Mussot2008,leo2013}. It is important to point out that in the absence of the filter the CW solution is modulationally stable, for the resonator parameters considered here, and thus no sidebands can be generated.

The relatively small number of comb lines observed is due to the low finesse of our resonator ($F=12$); high finesse resonators would enable a higher efficiency of the GTF process hence supporting a spectrally broader comb.
\newline

\begin{figure}[ht!]
\centering
\includegraphics[width=10cm,keepaspectratio]{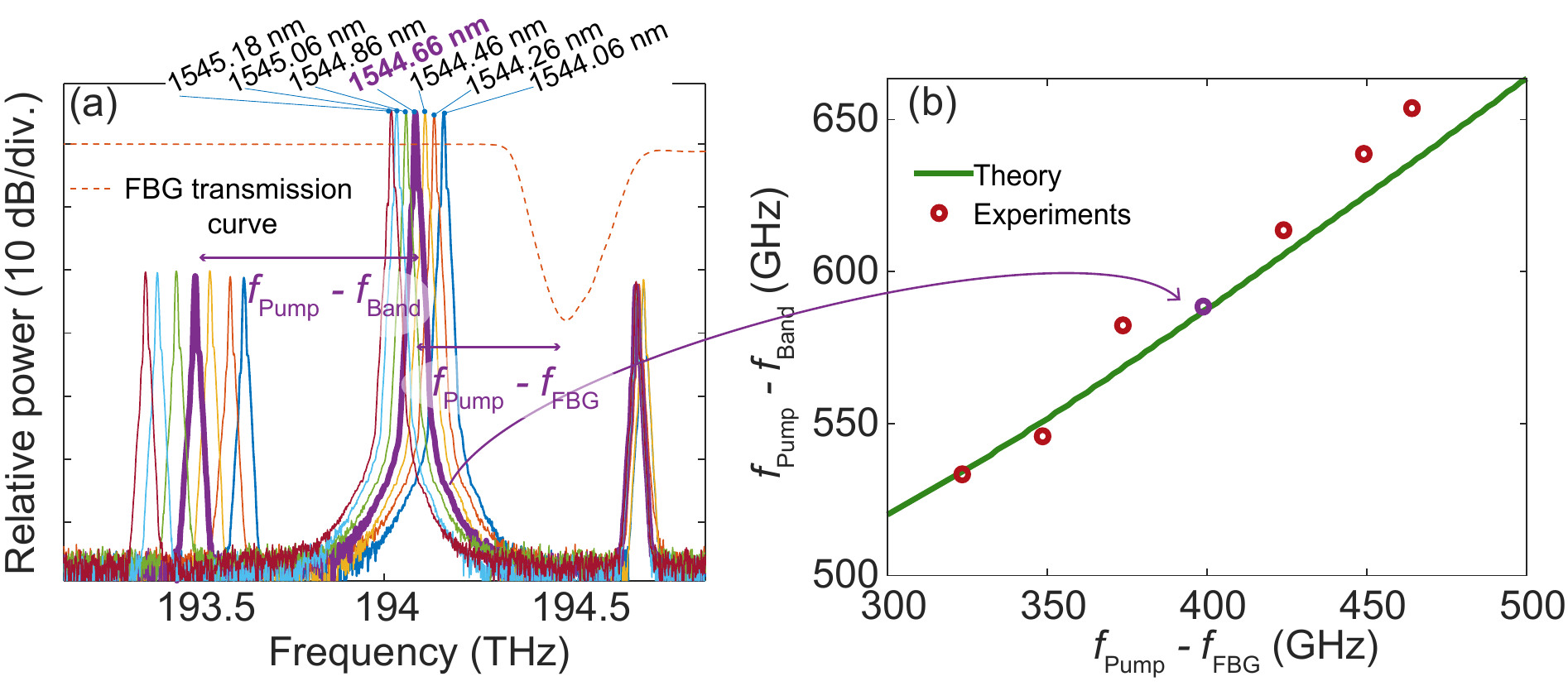}
\caption{Comb tuneability. (a) Output spectra for different pump wavelengths (from 1544.06 nm to 1545.18 nm). (b) Measured repetition rate as a function of the pump to fibre Bragg grating frequency shift (dots). The violet dot corresponds to the output spectrum depicted in violet in (a). The green line represents the frequency which maximises the instability gain calculated from Eq. (\ref{gain}) in Methods.}
\label{fig:3}
\end{figure}

\begin{figure}[ht!]
\centering
\includegraphics[width=10cm,keepaspectratio]{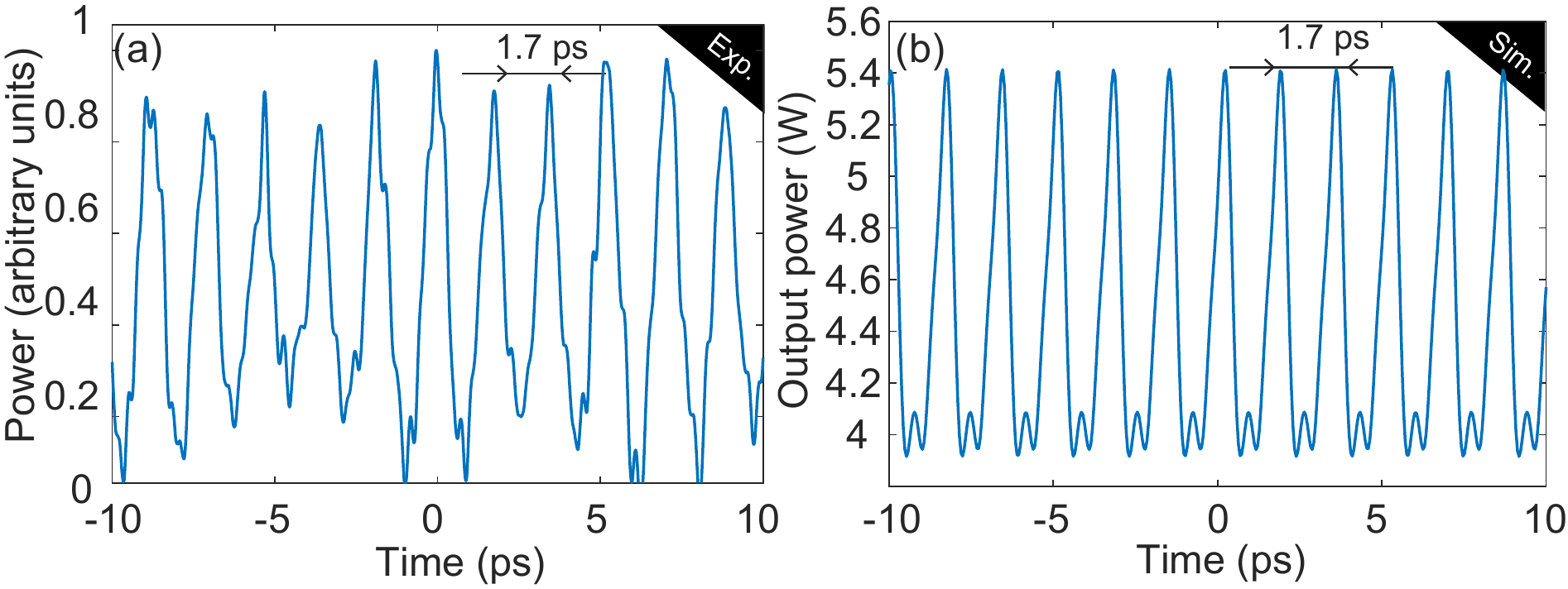}
\caption{Temporal dynamics. 
Temporal traces recorded at the output of the cavity show good agreement between experiments (a) and numerical simulations (b). Note that in experiment the background was attenuated before sending the optical signal to the oscilloscope, this explains the arbitrary units on the y-axis.}
\label{fig:4}
\end{figure}

\textbf{Comb tuneability.} In Fig. 3, we illustrate the tuneability of the GTF process. Keeping the central frequency of the filter unchanged, the pump wavelength is tuned from 1544.06 nm to 1545.18 nm leading to a frequency shift from the position of maximum loss of the filter from 323 GHz to 464 GHz. As seen from Fig. \ref{fig:3} (a), the position of the first anti-Stokes sideband remains in a fixed position, which is set by the filter position.
The corresponding Stokes sideband is symmetrically generated on the other side of the pump. For the sake of clarity, a close-up inspection is performed on the main three waves of the spectrum. It is important to stress that all spectra, whose zoom is depicted in Fig. \ref{fig:3} (a), are very similar to the frequency comb shown in Fig. \ref{fig:2} (a) corresponding to $\lambda_\text{P}=1544.66$ nm (bold violet curve). The measured repetition rate as a function of the frequency shift between the pump and the filter is shown in Fig. \ref{fig:3} (b)  (red dots), and is in excellent agreement with the theoretical results from linear stability analysis (green curve obtained from Eq. (\ref{gain}) in Methods). 
The comb repetition rate differs from the frequency shift of the pump from the filter, since as mentioned above the GTF sidebands are shifted from the maximum loss frequency of the filter. Consequently, by simply tuning the pump wavelength, we have demonstrated that the repetition rate of the frequency comb generated can be tuned from 533 GHz to 653 GHz in this particular case. 
\newline\newline
\textbf{Time domain waveform.} A frequency comb consists in phase locked equally spaced laser lines. Different complex experimental measurements can be performed to characterise the phase locking between the combs teeth. As a first  marker, the temporal trace can provide a very good insight into the spectral lines' coherence. As our frequency comb repetition rate is well above the bandpass of conventional electronic devices, we implemented a time lens system at the cavity output \cite{gaeta}. In this ways, we magnified the bandpass of our detection system (oscilloscope and photodetector) about 57 times, from 70 GHz up to 4 THz, to reach the repetition rate of the frequency comb. One example of temporal trace is represented in Fig. \ref{fig:4} (a). Localised temporal structured with a period of $\simeq$1.7 ps can be observed, that is indeed almost equal to 1/588 GHz. This is in a good agreement with numerical simulations results shown in Fig. \ref{fig:4} (b), where a period of 1.7 ps is found too. The fact that localised periodic structures can be observed in the time domain confirms that the comb lines are indeed phase locked. By measuring the temporal trace several times with a time separation of several seconds between consecutive measurements, a pattern with the same periodicity has been obtained, demonstrating the good stability of the generated structures.

At variance with both bright\cite{Kipp2} and dark \cite{Dark, Parra} cavity solitons, which require bistable response of the resonator, and hence non zero detuning of the pump wavelength with respect to the closest cavity resonance, in our case bright pulses, arising from amplified GTF modulation instability, exist at zero cavity detuning, and they do not require bistability.
\newline\newline\newline
 \textbf{\large Discussion}
 
 The major advantage of GTF as comb formation mechanism compared to existing methods is twofold: selectivity and tuneability. Indeed GTF allows selective amplification due to its operation in the normal dispersion regime. It is worth mentioning the fact that the background noise is not amplified for frequencies that do not satisfy the relation $f\approx f_\text{P}\pm n f_\text{g}$; where $f_\text{P}$ is the pump frequency, $f_\text{g}=\omega_\text{g}/(2\pi)$ with $\omega_\text{g}$ being the angular frequency satisfying the phase-matching condition (Eq.\ref{PM}), and $n$ is an integer. This feature is at variance with standard parametric amplification where the gain is continuously distributed from the pump up to a certain cut-off frequency hence amplifying without distinction all the resonator modes in such range, and can be exploited to achieve OFCs generation with better signal-to-noise ratio. Concerning tuneability, the comb repetition rate can be varied easily independently on pump power and in an all-optical fashion, by simply adjusting the pump-filter frequency detuning, without requiring complicated electronic devices \cite{Yao,quadTun}. This guarantees the transition from fundamental to potentially arbitrary harmonic comb. 
 While tuneability of OFCs line spacing has been demonstrated in Kerr resonators with tuneable pump wavelength \cite{DelHaye2}, in such case anomalous dispersion was needed since OFCs generation required standard parametric amplification process to take place. On the contrary our mechanism is more flexible, it does not require any dispersion engineering and is hence also very promising for OFCs generation in the visible part of the spectrum where material dispersion is normal.

In conclusion, we have proposed and demonstrated experimentally the first tuneable OFCs generation through filter induced MI in the normal dispersion regime. Our work introduces a paradigm-changing technique for OFCs generation in driven optical resonators that can be engineered and implemented in miniaturised systems, hence providing to end-users frequency combs with controllable features. 
\newline \newline

\textbf{\large Methods}
\newline \newline
\textbf{Experiments} We employed a tunable laser arround 1545 nm delivering square shaped pulses of 1.5 ns duration. The cavity is built with 104.2 m of a specially designed dispersion shifted fibre ($\beta_2^\text{DSF}$ at the pump wavelength is $0.5$ ps$^2$/km), a 90/10 coupler from the same fiber and a fiber Bragg grating used in transmission acting as a thin notch filter of 330 GHz at full width at half maximum centered at 1541.495 nm. The cavity is stabilized by a feedback loop system (proportional-intergral-derivative-controller) to keep both the driving frequency of the tuneable laser and the overall cavity length stable. Cavity output is recorded by an optical spectrum analyzer and with a commercial time lens system (Thorlabs) with a magnification factor of 57 allowing to get an effective $300$ fs resolution by using a set of fast oscilloscope and photodetector (70 GHz bandpass each). 
\newline\newline
\textbf{Numerics} The numerical simulations have been performed by solving a generalised Ikeda map.
The propagation of the electric field envelope $A(z,t)$, defined in the local time reference frame $t$ and evolving along the spatial coordinate $z$, through the ring fibre resonator during the roundtrip $n$ is described by the following nonlinear Schr\"{o}dinger equation:
\begin{eqnarray}
\frac{\partial A_n}{\partial z}=-i\frac{\beta_2}{2}\frac{\partial^2A_n}{\partial t^2}+\frac{\beta_3}{6}\frac{\partial^3A_n}{\partial t^3}+i\gamma|A_n|^2A_n.
\end{eqnarray}
Here $\beta_2$ and $\beta_3$ denote second and third order dispersion respectively, $\gamma$ is the nonlinearity coefficient, $0<z<L$ where $L$ is the fibre length. 
The propagation equation is coupled to the following boundary conditions, lumping together the effects of pumping, loss and filtering:

\begin{eqnarray}
A_{n+1}(z=0,t)=\theta E_\text{IN} +\rho e^{i\phi_0}h(t)* A_{n}(z=L,t).
\end{eqnarray}
Here $P_\text{IN}=|E_\text{IN}|^2$ is the power of the injected pump, $\theta$ is the coupler strength and $\phi_\text{0}$ is the linear phase shift between the pump and the intracavity field after one roundtrip. All the losses (propagation, splices,...) have been accounted for in the reflection coefficient $\rho$, so that $1-\rho^2$ measures the total power loss per roundtrip. $h(t)$ is the filter impulse response, which is assumed to be  causal ($h(t)=0$ for $t<0$) and the asterisk denotes convolution.

The complex filter transfer function is  the Fourier transform of the impulse response $H(\omega)=\mathcal{F}[h(t)]=\exp[\alpha(\omega)+i\psi(\omega)]$. The modulus of the filter $|H(\omega)|=\exp[\alpha(\omega)]$ is obtained from the measured filter attenuation profile, whereas the filter phase is computed from the Bode's magnitude-phase relation: $\psi(\omega)=\arg[H(\omega)]=-\mathcal{H}[\alpha(\omega)]$ \cite{bode}, being $\mathcal{H}[.]$ the Hilbert transform \cite{KK}. The presence of the filter phase ensures that Kramers-Kronig relation and hence causality is preserved. We have defined the magnitude of the filter as: $|H(\omega)|=1-R\exp[-(\omega-\omega_\text{f})^4/(\sigma_\text{f})^4)]$, which gives a quite accurate approximation of the filter used in the experiments. 
The cavity output is given by the following relation, which accounts for the phase introduced by the fibre coupler and the accumulated total loss \cite{stokes}
\begin{eqnarray}
A_{\text{OUT},n}=i\sqrt{1-\theta^2}E_\text{IN}-i\eta\theta e^{i\phi_0} A_{n}(z=L,t),
\end{eqnarray}
where  $\eta=\rho/\sqrt{1-\theta^2}<1$.
The map is integrated for a sufficiently high number of roundtrips in order to reach a stationary state.
The nonlinear Schr\"{o}dinger equation has been numerically integrated using a standard split-step Fourier method and the following set of parameters has been used: $\beta_2=0.5$ ps$^2$/km , $\beta_3= 0.12$ ps$^3$/km, $\gamma=2.5$ W$^{-1}$km$^{-1}$, $L=104.2$ m, $\rho^2=0.5940$ (cavity finesse $F=12$) , $\theta^2=0.1$, $\phi_0=-\arg[H(0)]$ (the measured total cavity detuning is zero), $R=0.96$, $\omega_\text{f}=399\times 2\pi$ rad/ns (in Figs. \ref{fig:2} and \ref{fig:4})
, $\sigma_\text{f}=160 \times 2\pi$ rad/ns, $P_\text{IN}=6.6$ W.
\newline

\textbf{Theory}
Linear stability analysis allows to calculate the frequency and the gain of the spectral components amplified by the GTF mechanism. The full calculations are reported in the Supplementary Information (see Supplementary Note 3). We report here only the final result. The amplitudes of unstable spectral components grow from noise as $\exp[g(\omega)z]$, where the gain is 
\begin{equation}\label{gain}
g(\omega)=\frac{1}{L}\ln\left(\max\left|\frac{\Delta}{2}\pm\sqrt{\frac{\Delta^2}{4}-W}\right|\right),
\end{equation}
and $ W=\rho^2(H_\text{e}(\omega)^2+H_\text{o}(\omega)^2)$, $\Delta=\rho[2\cos(kL)(H_\text{e}(\omega) \cos \phi - H_\text{o}(\omega) \sin \phi)- \sin(kL)(H_\text{o}(\omega) \cos \phi +H_\text{e}(\omega) \sin \phi)(\beta_2\omega^2+2\gamma P)/k ]$,
$H_\text{e,o}(\omega)=(H(\omega)\pm H(-\omega)^*)/2$ are the even and odd part of the filter transfer function $H(\omega)$, and $k=\sqrt{\beta_2\omega^2/2(\beta_2\omega^2/2+\gamma P)}$.

As detailed in the Supplementary Information, the frequency which maximises the gain Eq.(\ref{gain}) is well approximated by the solution of the phase-matching Eq.(\ref{PM}).
In the Supplementary Figure 3 the dependence of the gain on various parameters is graphically depicted too.
\newline

\textbf{Data availability}\newline
Data are available from the corresponding author upon reasonable request.
\newpage\noindent

\textbf{Acknowledgments}
\item We acknowledge Laure Lago for fabricating FBGs used in this experiment. This work was partly supported by the Agence Nationale de la Recherche through the Labex Centre Europeen pour les Mathematiques, la Physique et leurs Interactions (CEMPI) and Equipex Fibres optiques pour les hauts flux (FLUX) through the ``Programme Investissements d'Avenir", by the Ministry of Higher Education and Research, Hauts de France council and European Regional Development Fund (ERDF) through the Contrat de Projets Etat-Region (CPER Photonics for Society, P4S) and FEDER through the HEAFISY project. K.S. acknowledges support from Leverhulme Trust Fellowship.

\item[\textbf{Authors contributions}] A.M.P., K.S. and S.K.T. initiated the study;
F.B., A.M., A.K. and M.C. conceived the experimental setup. F.B., A.M. and A.K. performed the experiment.  A.M.P. and M.C. performed the numerical simulations and developed the theory. All authors contributed to analysing the data and
writing the paper.

\item[\textbf{Competing Interests}] The authors declare no competing interests.
\item[\textbf{Correspondence}] Correspondence and requests for materials should be addressed to A. M. Perego~(email: a.perego1@aston.ac.uk). Most of the relevant data used in this paper are contained in the Supplementary Information while further data are available from the corresponding authors upon reasonable request.

\end{document}